Statistics of dislocation pinning at localized obstacles

A. Dutta[a], M. Bhattacharya[b,*], and P. Barat[b]

[a]S. N. Bose National Centre for Basic Sciences, JD Block, Kolkata 700 098, India

[b]Variable Energy Cyclotron Centre, 1/AF, Bidhannagar, Kolkata 700 064, India

**Abstract**

Pinning of dislocations at nanosized obstacles like precipitates, voids and bubbles, is a crucial mechanism in the context of phenomena like hardening and creep. The interaction between such an obstacle and a dislocation is often explored at fundamental level by means of analytical tools, atomistic simulations and finite element methods. Nevertheless, the information extracted from such studies has not been utilized to its maximum extent on account of insufficient information about the underlying statistics of this process comprising a large number of dislocations and obstacles in a system. Here we propose a new statistical approach, where the statistics of pinning of dislocations by idealized spherical obstacles is explored by taking into account the generalized size-distribution of the obstacles along with the dislocation density within a three-dimensional framework. The application of this approach, in combination with the knowledge of fundamental dislocation-obstacle interactions, has successfully been demonstrated for dislocation pinning at nanovoids in neutron irradiated type 316-stainless steel in regard to both conservative and non-conservative motions of dislocations.

*Corresponding author:

Dr. Mishreyee Bhattacharya
Email: mishreyee@vecc.gov.in

## 1. Introduction

Nanosized defects like precipitates, voids and bubbles can act as strong localized obstacles by pinning the glide or climb motion of dislocations, thereby dictating the mechanical behaviour of a metallic solid over wide range of length and time scales. A large number of studies have already been performed over the last few decades to reveal the nature of interactions between dislocations and localized obstacles. Such investigations cover multiple domains of experiments, analytical modelling and computer simulations. As the outcome of such efforts, sufficient understanding of the phenomenon of dislocation pinning at nanoscale defects has been gained. For example, in the context of dislocation-precipitate interactions, several strengthening mechanisms like incoherency strengthening [1-2], stacking-fault resistance [3], atomic order strengthening [4], misfits in size and elastic modulus [5] etc., are understood qualitatively and to a satisfactory extent, quantitatively as well. Similarly, the role of defects like voids and bubbles has also been examined not only in terms of their effectiveness in offering resistance against the glide of dislocations [6-9], but also in governing the climb motion via diffusive [10] and non-diffusive [11] routes.

Despite the significant progress in understanding the basic mechanisms of dislocation pinning at different types of localized obstacles, the capability of predicting their effects on the gross behaviour of materials at macroscopic scales has remained elusive. Any attempt to tackle this problem by extrapolating the fundamental dislocation-obstacle interactions to a multiscale domain invariably suffers from the lack of information regarding the statistics of pinning of dislocations at these obstacles. Significant attempts have been made in this direction in the form of Friedel-Fleischer [12,13] and Mott-Labusch [14,15] statistics from 1950's to 1970's. The analytical results were extensively compared with the computer simulations of gliding dislocations on a given plane and the ranges of operational and material parameters over which the theoretical estimates held their validity were identified.

While deriving the flow stresses with simplest assumptions, all the obstacles in the medium were supposed to be of the same size, i.e., the mean value of the entire size distribution. For the mixtures of two types of obstacles, Kocks *et al*. [16] suggested a linear sum of the flow stresses while Koppenaal and Kuhlmann-Wilsdorf [17] introduced the so called Pythagorean sum, the predictions from which were found to be closer to the simulation. Such studies dealt with the sampling statistics of a dislocation line encountering an obstacle during its motion on a two-dimensional domain representing a given glide plane. Nonetheless, any statistical framework is still awaited, which can take into account both the dislocation density of the medium and the complete size-distribution of the obstacles in three dimensions to yield the detailed statistics of pinning across the entire size-range of the obstacles. The knowledge of pinning-statistics becomes more relevant in view of the fact that the observable parameters of practical importance often exhibit highly nonlinear relation with the size of the obstacle or the dislocation link-length. For example, in the case of hardening induced by nanosized voids in both b.c.c. and f.c.c. solids, the critical resolved shear stress (CRSS) required to depin a dislocation line follows a complex logarithmic relation with the size of the void [6-8]. In the same way, the amount of shrinkage experienced by a nanovoid during the void-enhanced climb of a pinned dislocation line [10] exhibits a nonlinear relation with the original void size. Hence, it is clear that the method of simple averaging essentially obscures the valuable information regarding the collective dynamics of the entire dislocation-obstacle system, which could have been extracted, had the actual statistics of pinning been available.

In this paper, we propose a new approach to deal with the above mentioned issue. With the help of simple geometrical arguments and a few justified assumptions, we derive a closed form solution for the statistics of dislocation pinning at localized nanosized obstacles of arbitrary size-distribution. This treatment is expected to hold for homogeneous distributions of dislocations and spherical obstacles. As a case study, we demonstrate the applicability of our results by considering the formation of nanovoids due to radiation damage in type 316-stainless steel. The statistical framework has been employed to understand the effect of these voids on the climb and glide motions of a large number of pinned dislocations. The analyses show that by means of the statistics derived here, it is possible to uncover some important aspects of the coarse and collective behaviour of the system from the existing knowledge of dislocation-obstacle interactions.

## 2. Theoretical framework

This section begins with the description of the model and discusses the validity of the basic assumptions in light of the realistic systems. Later on, we present a geometric approach, which leads to the final solution of the statistical problem considered here.

### 2.1 *Origin of the model*

We consider a unit cube (Fig. 1) of the material such that the sides of the cube are of macroscopic order and the following statistical treatment can be applied on this domain. We assume a homogeneous distribution of spherical obstacles and a three-dimensional network of dislocations. At this point, we do not distinguish among the types of dislocations (screw or edge, orientation, glide plane etc.) or the obstacles (voids, bubbles or precipitates). However, it will later be argued that the final results are equally applicable to any given combination of dislocations and obstacles, provided that their numbers are large enough to fit the statistical validity. Our treatment is aimed at developing a final general solution of the pinning statistics, which would depend only on the small set of material parameters listed below:

$\rho_{\perp}$ : Density of dislocations in the material,

$N_o$ : Number of spherical obstacles in the unit cube (number density),

$\eta_o$ : Volume fraction of the obstacles,

$f_o(r)$ : Obstacle size-distribution, i.e., the probability density of the obstacle of radius $r$.

By definition, the obstacle size distribution follows the normalization criterion,

$$\int_{r_{min}}^{r_{max}} f_o(r)dr = 1, \qquad (1)$$

where $r_{min}$ and $r_{max}$ denote the lower and upper bounds of the size of obstacles, respectively. The above normalization relation yields the mean volume of the obstacles as,

$$\langle V_o \rangle = \frac{4}{3}\pi \int_{r_{min}}^{r_{max}} r^3 f_o(r)dr. \qquad (2)$$

Therefore, the number density of the obstacles can be written as,

$$N_o = \frac{\eta_o}{\langle V_o \rangle} = \frac{3\eta_o}{4\pi \int_{r_{min}}^{r_{max}} r^3 f_o(r)dr} . \qquad (3)$$

2.2 *Meshing statistics*

We consider all the obstacles of an arbitrary radius, $r$, in the cube. Each of these obstacles can be imagined to be circumscribed by a cubical cell of side $2r$ (Fig. 2a). The volume of such a cell is given as $V_{\boxdot}(r) = 8r^3$ and the number of such cells, which are required to fill the entire volume of the unit cube, can be expressed as,

$$N_{\boxdot}(r) = V_{\boxdot}(r)^{-1} = \frac{1}{8r^3} . \qquad (4)$$

Thus, we can partition the unit cube sample into $N_{\boxdot}(r)$ cubic meshes of sides $2r$ (Fig. 2b). For small dislocation density, only a fraction of these meshes would contain a part of the dislocation network, while the rest of them would be free of these line defects. However, as the dislocation density increases, each of the meshes would have to accommodate some part of the dislocation network. We assume that all the cells containing the dislocations, form an ensemble of, say $N_{\perp}(r)$ meshes, such that $N_{\perp}(r) \ll N_{\boxdot}(r)$ for very low dislocation densities whereas $N_{\perp}(r) \rightarrow N_{\boxdot}(r)$ as the density increases. Each of the dislocation-containing cells possesses a small dislocation segment originating from one of the faces and ending up on any of the remaining five as shown schematically in Fig. 2b. As the cubic cells are nanosized and curvature of the dislocation line cannot be too large on account of its large line tension, which is on the order of $Gb^2/2$ ($G$ is the shear modulus and $b$ is the Burgers vector), one can reasonably assume that the dislocation line is piecewise-straight within a given cell. Therefore, using Eq. (4), the fraction of cells which contain parts of the three-dimensional dislocation network becomes,

$$F_{\perp}(r) = \frac{N_{\perp}(r)}{N_{\boxdot}(r)} = 8r^3 N_{\perp}(r) \,. \qquad (5)$$

If $\langle L_{\perp}(r)\rangle$ is the mean length of the dislocation segment over the $N_{\perp}(r)$ dislocation-containing meshes, we have by definition,

$$N_{\perp}(r) = \frac{\rho_{\perp}}{\langle L_{\perp}(r)\rangle} \,. \qquad (6)$$

By substituting Eq. (6) into Eq. (5), one arrives at the relation,

$$F_{\perp}(r) = \frac{8r^3 \rho_{\perp}}{\langle L_{\perp}(r)\rangle} \,. \qquad (7)$$

The above equation shows that the fraction $F_{\perp}(r) < 1$ for small dislocation densities, but exceeds unity as $\rho_{\perp} > \frac{\langle L_{\perp}(r)\rangle}{8r^3}$. Here it must be pointed out that the possibility of $F_{\perp}(r)$ becoming larger than unity in accordance with Eq. (7) is not an aberration in conflict with Eq. (5). It simply implies a physical situation, where the dislocation density is so large that a single cubic mesh packs multiple dislocation segments. Accordingly, a cubic cell containing more than one dislocation segment is counted multiple times to effectively render $N_{\perp}(r) > N_{\boxdot}(r)$.

When the entire medium is partitioned into numerous cubic meshes of side $2r$, most of the obstacles of radius $r$ would not have their centres coinciding with the centre of any cell. Here we can assume that all such obstacles are given small displacements, so that they can properly be inscribed in their nearest cells as shown schematically in Fig. 3. This process can be repeated for all the possible values of obstacle-radius, $r$. Although, these shifts change the microscopic details of the structure on a fine spatial scale, they do not alter the coarse structural features and statistics of dislocation pinning, provided that the number of obstacles are large enough in the unit volume of the solid. This is because, if such a displacement causes an obstacle to release a previously pinned dislocation line with a given probability, there would be another identical obstacle, which would now get displaced to pin a dislocation line with an equal statistical likelihood. In other words, the superposition of small random perturbations on an initial random spatial distribution does not affect the measure of randomness of the final results. Once all the obstacles are fitted in their respective cells, we have the number density of cells with obstacles of radius in the narrow range [$r$, $r + dr$] as equal to the number of obstacles with that size in the sample. This can be written using Eq. (3) as

$$\rho_{\mathrm{o}}(r)dr = N_{\mathrm{o}} f_o(r)dr = \frac{3\eta_{\mathrm{o}} f_o(r)dr}{4\pi \int_{r_{min}}^{r_{max}} r^3 f_o(r)dr} . \qquad (8)$$

Therefore, using Eqs. (4) and (8), the fraction of cells with obstacles in the selected size-range [$r$, $r + dr$] is obtained as,

$$F_{\mathrm{o}}(r)dr = \frac{\rho_{\mathrm{o}}(r)dr}{N_{\square}(r)} = \frac{6\eta_{\mathrm{o}} r^3 f_o(r)dr}{\pi \int_{r_{min}}^{r_{max}} r^3 f_o(r)dr}. \qquad (9)$$

2.3 *Pinning statistics*

In our statistical model, we define pinning as the coexistence of an obstacle and part of a dislocation line in the same cell of the meshed sample. It can be pointed out that if the dislocation is regarded as a geometrical curve or line with zero lateral dimension, it is possible that a part of the dislocation line can be accommodated in the cell without touching the obstacle. However, as we have assumed obstacles to be nanosized and the real dislocations may have core widths spanning up to several Burgers vector, the dislocation-obstacle interaction would be strong enough if both of those belong to the same cell and an effective pinning can be assumed for all practical purposes. Hence, from Eqs. (7) and (9), the probability-density that a given cell accommodates both a dislocation segment and an obstacle simultaneously, can be written as,

$$F_{\perp o}(r) = F_{\perp}(r)F_{o}(r) = \frac{48\eta_o \rho_{\perp} r^6 f_o(r)}{\pi\langle L_{\perp}(r)\rangle \int_{r_{min}}^{r_{max}} r^3 f_o(r)dr}\,. \qquad (10)$$

So far, the only unknown term is $\langle L_{\perp}(r)\rangle$, which is the mean length of dislocation segments in the small cubical meshes. As the dislocation segment within a cell is assumed to be piecewise-straight, evaluating $\langle L_{\perp}(r)\rangle$ is reduced to the problem of obtaining the expected value of the distance between two random points selected on different faces of a cube, which belongs to the class of the so called 'line-picking' problems. One can note that the present treatment does not take the dislocation junctions into account. However, as the meshes are very small in size, the total number of straight dislocation segments is much larger than that of dislocation junctions and the perturbing effect on the estimated value of $\langle L_{\perp}(r)\rangle$ is negligibly small. Statistical line-picking problems were first attempted by Ghosh [18-21] for rectangular domains, whereas the solutions for their 3-dimensional counterparts are more recent. The mean distance between two points on different faces of a unit cube (Fig. 4) is given by [22],

$$\Delta_f(3) = \frac{4}{5}\int_0^1\int_0^1\int_0^1\int_0^1 \{x^2 + y^2 + (z-w)^2\} dwdxdydz + \frac{1}{5}\int_0^1\int_0^1\int_0^1\int_0^1 \{1 + (y-u)^2 + (z-w)^2\} dudwdydz\,. \qquad (11)$$

The above equation is composed of two weighted terms, where the first term represents the pairs of points on adjacent faces, while the second one represents those on the opposite faces of a unit cube. These expressions have been solved analytically and experimentally to yield the final result as [22],

$$\Delta_f(3) = \frac{7}{5}\Delta(3) = \frac{1}{75}\left\{4 + 17\sqrt{2} - 6\sqrt{3} + 21\ln\left(1+\sqrt{2}\right) + 42\ln\left(2+\sqrt{3}\right) - 7\pi\right\}, \quad (12)$$

where $\Delta(3)$, also known as the Robbins constant [23], denotes the mean distance between two random points in the interior of the unit cube. $\Delta_f(3)$ is evaluated as 0.926 and thus, we have $\langle L_\perp(r)\rangle = 2r \times 0.926 = 1.852r$. On substituting this value in Eq. (10), we obtain,

$$F_{\perp\mathrm{o}}(r) = \frac{8.25\eta_{\mathrm{o}}\rho_\perp r^5 f_o(r)}{\int_{r_{min}}^{r_{max}} r^3 f_o(r)dr}. \qquad (13)$$

If $N_{\perp\mathrm{o}}(r)$ is the number of pinning sites at obstacles with radii in the range [$R_1$, $R_2$] in the unit volume, from the above probability-density function and Eq. (4), we get

$$N_{\perp\mathrm{o}}(R_1; R_2) = \int_{R_1}^{R_2} N_{\boxdot}(r)F_{\perp\mathrm{o}}(r)dr = \frac{1.03\eta_{\mathrm{o}}\rho_\perp \int_{R_1}^{R_2} r^2 f_o(r)dr}{\int_{r_{min}}^{r_{max}} r^3 f_o(r)dr}, \qquad (14)$$

which can be rewritten in terms of the size distribution as

$$\frac{dN_{\perp\mathrm{o}}}{dr} = \frac{1.03\eta_{\mathrm{o}}\rho_\perp r^2 f_o(r)}{\int_{r_{min}}^{r_{max}} r^3 f_o(r)dr}. \qquad (15)$$

Equation (15) demonstrates that the complex spatial statistics of dislocation pinning at nanosized spherical obstacles is ultimately reducible to a simple and rigid expression involving only the basic parameters of the material under consideration. So far, we have considered the term $F_{\perp\mathrm{o}}(r)$ in Eqs. (10) and (13) as the probability density and assumed that a single cubic mesh cannot contain more than one segment of the dislocation line. However, having obtained Eq. (15), one can realize that this assumption is not a rigid constraint. It simply means that we can have $\frac{dN_{\perp\mathrm{o}}}{dr} > \frac{dN_{\mathrm{o}}}{dr}$ at large dislocation densities. Such a situation will arise when $\rho_\perp$ exceeds its critical value $\left(\frac{\langle L_\perp(r)\rangle}{8r^3}\right)$ for transition from light to dense packing, so that $F_\perp(r) > 1$ in accordance with Eq. (7). In that case, $N_{\perp\mathrm{o}}(r)$, as computed from Eq. (14) or (15) would represent the number of dislocation segments that get pinned by obstacles of a given size-range. Obviously, this number of pinned dislocation segments will be larger than that of the obstacles themselves in case of multiple pinning at the same defect site as mentioned earlier.

## 3. Case study: pinning statistics for nanovoids in irradiated type 316-stainless steel

In nuclear structural materials, the fast-neutron fluence is well known to create very large number of vacancies and interstitials due to radiation damage. At moderate rate of point-defect diffusion, the interstitials agglomerate to form interstitial dislocation loops, whereas the vacancies may create either vacancy-loops or nanosized voids with higher propensity to the latter one [24]. In the present section, we employ the statistical framework developed in the previous section and compute the statistics of dislocation pinning at the nanovoids formed at different temperatures. Moreover, this information is further used to analyze the influence of pinning on the non-conservative (climb) and conservative (glide) motions of dislocations.

### 3.1 *Crossover from rare pinning to multiple pinning*

As a model system, we consider the radiation damage of type 316-stainless steel at the neutron fluence of $6\times10^{22}$ neutrons/cm$^2$. At this fluence, the size distributions of the nanovoids ($f_o(r)$) are extracted for temperatures 653 K, 713 K, 763 K, 803 K and 853 K, from the results given in Fig. 19.1 of Ref. [24], while the volume fraction $\eta_o$ (Table-1) is obtained from an empirical relation [24,25]. As the formation of nanovoids and dislocation loops are in competition with each other, we can assume that the dislocation loops and the voids are spatially separated. Therefore, we emphasize only on the pinning of network dislocations. The densities ($\rho_\perp$) of these line defects are calculated (Table-1) using the empirical relation given by Brager and Straalsund [24,26] for the type 316-stainless steel. Having all the parameters required for Eq. (15) at hand, we can now compute the distributions of dislocation-nanovoid pinning at different temperatures.

Figures 5a-e show the statistics of pinning along with the original void-size distributions at the five different temperatures specified above. At 653 K, the range ($r_{max}$ -

$r_{min}$) of the size distribution is narrow and the network dislocation density is also very small, i.e., ~$10^8$ cm$^{-2}$ (Table-1). As a result, the likelihood of dislocation-obstacle interaction is small and $\frac{dN_{\perp o}}{dr} \ll \frac{dN_o}{dr}$ (rare pinning), over the entire range of void radius as displayed in Fig. 5a. At 713 K (Fig. 5b), the size range of the nanovoids is somewhat broader and the dislocation density increases by about two orders of magnitude (Table-1). At this temperature, each of the voids near the extreme upper end of the size-distribution is large enough to capture a single dislocation segment on the average and the plots for $\frac{dN_{\perp o}}{dr}$ and $\frac{dN_o}{dr}$ merge into each other. A crossover is noticeable for 763 K (Fig. 5c), where both dislocation density and the average void-size are large. As mentioned in Sec. 2, void-sizes above this crossover value indicate multiple pinning, where a single obstacle pins more than one dislocation segment. At even higher temperatures of 803 K and 853 K, the void-sizes increase drastically while a high dislocation density is maintained (refer Table-1). Consequently, most of the obstacles entertain multiple pinning and the crossover point eventually shifts towards the lower end of the size-distribution as evident from Figs. 5d and e. Apart from the rigorous statistical treatment presented in the previous section, it is possible to understand the physical significance of crossovers between the $\frac{dN_{\perp o}}{dr}$ and $\frac{dN_o}{dr}$ plots from a more heuristic perspective. For the dislocation density $\rho_\perp$, the mean distance between two adjacent dislocation lines is approximated as $\rho_\perp^{-1/2}$. If an obstacle diameter exceeds this mean distance, it would pin multiple segments simultaneously. It can be observed that the crossover radii for the plots in Figs. 5b-e indeed lie close to $\rho_\perp^{-1/2}/2$, thereby validating the consistency of Eq. (15).

3.2 *Void-assisted diffusive climb of dislocations*

Knight and Burton [10] proposed a mechanism and model for the recovery process of void-swelling in irradiated materials. In this model, the voids connected to dislocations shrink in size by pumping vacancies into the cores of the dislocations, thereby causing them to undergo the climb motion. Unlike the case of non-diffusive climb as revealed in our earlier study [11], this mechanism involves pipe-diffusion of vacancies and is a very slow process. We start with the idealized model outlined by Knight and Burton [10], where a single dislocation segment connects two voids and maintains a curvature due to climb as represented schematically in Fig. 6. On account of the inherent line-tension of the dislocation segment, a void experiences a drag and moves a distance $y$ due to its mobility controlled by lattice and surface diffusions [27]. Accordingly, the area swept by the dislocation segment during the climb is given as

$$A = \lambda y + \left[R_c^2 sin^{-1}\left(\frac{\lambda}{2R_c}\right) - \left(\frac{\lambda}{2}\right)^2 \left\{\left(\frac{2R_c}{\lambda}\right)^2 - 1\right\}^{1/2}\right], \quad (16)$$

where $R_c$ is the radius of curvature of the dislocation link and $\lambda$ is the average link-length. In the above equation, the first term $\lambda y$ is the rectangular area corresponding to the drag of the voids over the distance $y$, and the second term belongs to the area swept due to the curvature of the dislocation (Fig. 6). In their original work [10], Knight and Burton considered the total area to be dominated by the first term only, an assumption which in the present study is valid for the irradiation temperatures of 653 K and 713 K, where the voids are small and multiple pinning at a single void is unlikely. Accordingly, the shrunk radius of a nanovoid of initial radius $r_0$ is given by [10],

$$r(y) = r_0 \left[1 - \frac{3b\lambda y}{4\pi r_0^3}\right]^{1/3}, \quad (17)$$

where the Burgers vector, $b$, can be taken as 0.25 nm. Even though, Eq. (17) encapsulated the fundamental features of the mechanism of void-shrinkage, it is still insufficient to yield the dynamically evolving configuration of the dislocation-void system. For example, we can observe that for a void of initial size $r_0$, the maximum limit of the displacement $y$ is $\frac{4\pi r_0^3}{3b\lambda}$, at which the void completely vanishes. However, as the voids pinning the dislocation lines follow a size distribution (Fig. 5), the smaller voids would vanish rapidly, followed by the larger ones. As soon as a fraction of the voids disappear, the effective link-length $\lambda$ would increase and induce a dynamic character to the form of Eq. (17). In the same way, continuous shrinkage and disappearance of the voids would cause the size distribution $\frac{dN_{\perp o}}{dr}$ and available range [$r_{min}$, $r_{max}$] to vary and accordingly, the volume fraction of the pinned voids, $\frac{4}{3}\pi \int_{r_{min}}^{r_{max}} r^3 \frac{dN_{\perp o}}{dr} dr$, becomes a function of the displacement $y$. As we have already been able to calculate the initial size distribution of the pinned nanovoids (Fig. 5), it becomes possible to compute the evolution of $\frac{dN_{\perp o}}{dr}$ using Eq. (17). For the present case, we can approximate the average link-length as,

$$\lambda(y) = \frac{\rho_{\perp}}{N_{\perp o}(y)}, \qquad (18)$$

where $N_{\perp o}(y)$ can be computed iteratively using Eqs. (15), (17) and (18). Figures 7a and b illustrate the rise in the link-length with displacements of the pinned nanovoids at 653 K and 713 K respectively. We find that in general, the initial link-length at 653 K is much larger than that at 713 K temperature. As the voids start disappearing due to loosing vacancies to the pinned dislocations by means of pipe diffusion, the effective link-lengths increase with the climb distance. In the same way, the volume fractions of the nanovoids (void-swelling) are shown in Figs. 7c and d for these two temperatures. One can observe that although a visible recovery is present at 713 K, it is negligibly small for the irradiation temperature of 653 K. The reason behind such a trend is clear from Fig. 5. At 653 K, only a tiny fraction of the total number of voids is pinned (Fig. 5a) and the overall volume fraction is almost completely dictated by the unpinned voids. Since the unpinned voids do not contribute to the mechanism of dislocation climb by core-diffusion of the vacancies, we observe its reflection on Fig. 7c accordingly. In contrast, a larger fraction of voids pin the network dislocations (Fig. 5b) at 713 K, so that we witness a more prominent recovery (Fig. 7d) as discussed above.

The modality of recovery is expected to be different at higher irradiation temperatures, where a single void has a high likelihood of pinning multiple dislocation lines due to larger size. Under such circumstance, the void would pump the vacancies into several dislocation cores simultaneously and shrink at a rate, which is much larger than that estimated by the simple idealized scheme considered so far [10]. Moreover, the dislocation segments would tend to bow-out in different directions and on different crystal planes and thus nullify the resultant drag on the pinning nanovoid. As a result, the void displacement $y$ in Eq. (16) can be ignored and we modify the total area swept by the curved dislocation lines as,

$$A(r;\ R_c) \approx\ \mu(r)\left[R_c^2 sin^{-1}\left(\frac{\lambda}{2R_c}\right) - \left(\frac{\lambda}{2}\right)^2\left\{\left(\frac{2R_c}{\lambda}\right)^2 - 1\right\}^{1/2}\right], \quad (19)$$

where the multiplicity factor, $\mu(r) = \frac{dN_{\perp 0}/dr}{dN_0/dr}$ (refer Figs. 5d and e) denotes the average number of times a single void of radius $r$ pins down the dislocation lines. It must be noted that this multiplicity factor is estimated only using the initial size distributions of the nanovoids, and remains invariant throughout the shrinkage. In the case of irradiation temperatures of 803 K and 853 K, $\mu(r)\ > 1$ for most of the nanovoids as evident from Figs. 5d and e. In such cases, the shrinkage of a given void of initial radius $r_0$ is expressible by modifying Eq. (17) as,

$$r(\theta) =\ r_0\left[1 - \frac{3bA}{4\pi r_0^3}\right]^{1/3}, \quad (20)$$

where the angle of curvature, $\theta$, relates to the radius of curvature as $\theta = sin^{-1}\frac{\lambda}{2R_c}$ (Fig. 6). Similar to the previous iterative computations for lower temperatures, we again estimate the evolution of link-length and volume recovery. One can notice that the relative increment of the link-length at the irradiation temperature of 853 K is much larger than that for 803 K (Figs. 8a and b). In the same way, the volume recovery is significantly high (Figs. 8c and d). The reasons behind such trends are twofold: first, the large multiplicity compels almost all the voids to contribute to the void-assisted dislocation climb at vigorous rates and second, the large sizes of the voids cause appreciable recovery without complete disappearance and consequent increase in the link-lengths. The situation is expected to be somewhat more complex for the case of irradiation at intermediate temperature of 763 K (Fig. 5c), where the small voids with low probabilities of pinning and high mobility would tend to behave according to Eq. (17), whereas the larger voids can follow the trend of Eqs. (19) and (20). Even though the void-shrinkage equations are still valid for this case, it may need a separate treatment if the entire size-distribution of the nanovoids is to be handled simultaneously. If the reliable data regarding the thermally controlled parameters like surface, lattice and pipe-diffusivities are available for the solid under consideration, Eqs. (17) and (20) can easily be extended [10] to yield the instantaneous recovery with respect to the elapsed time.

### 3.3 *Depinning at nanovoids under shear load*

Besides affecting the climb motion of dislocations, the radiation-induced formation of nanosized voids is also associated with the effect of void-hardening in the context of steady state glide motion of the dislocations. The nanovoids are known to act as strong pinning sites for the dislocation lines and exert substantial impedance to their glide mobility. A dislocation segment pinned at a nanovoid needs a critical resolved shear stress to get depinned at the obstacle. The mechanism of dislocation-nanovoid interaction has been studied earlier in detail at the fundamental level [6-9,11]. Extensive molecular dynamics simulations have shown that the CRSS required to depin a dislocation segment of link-length $\lambda$ at a nanovoid of radius $r$ is given by [8],

$$\tau_c = \frac{Gb}{2\pi\lambda}\left[ln(0.5r^{-1} + \lambda^{-1})^{-1} + 1.52\right]. \qquad (21)$$

Although the above expression has been specified for a particular glide plane on which the applied load is supposed to be resolved, we can assume an averaging pre-factor multiplied to Eq. (21), which could account for multiple slip-planes and directions. Even if this pre-factor is not available *a priori*, we can still estimate the effect on macroscopic yield stress due to void-hardening, provided that the stresses are expressed in normalized form. The molecular dynamics simulations, which were employed to test the validity of Eq. (21), were carried out with the hypothetical periodic boundary conditions. Consequently, as the depinning of the dislocation segment takes place in the primary simulation cell, it simultaneously occurs in all the image cells as well. Although such studies offer valuable insight of the hardening process at the local microscopic scale, we can further utilize such information to obtain a statistical picture of the glide of dislocations, using the formalism developed in the present study. In the realistic scenario, the smallest obstacles would release the pinned dislocation at comparatively smaller load. As soon as this happens, the effective mean link-length would increase, which in turn, would tend to alter the CRSS in accordance with Eq. (21). In order to explore the competition between the effects of increasing the link-length and void-size, we compute the evolution of CRSS in an iterative manner for all the irradiation temperatures specified in Sec. 3.1. We start with the nanovoids of smallest size and compute the corresponding CRSS using the instantaneous value of $\lambda = \frac{\rho_{\perp}}{N_{\perp o}}$. We assume that almost all the dislocation segments pinned at these smallest nanovoids get depinned at this applied shear load, thereby reducing the number of pinned segments, $N_{\perp o}$, and increasing the link-length accordingly. This modified value of $\lambda$ is then used to compute the CRSS for the next set of smallest voids pinning the dislocations. The typical evolution of the normalized CRSS is exhibited in Fig. 9 as the function of void-radius. It is informative to note that instead of the obstacles of largest size, the maximum CRSS value is obtained for the voids of much smaller radius. The reason behind this trend is the significant increment in the mean link-length, which nullifies and supersedes the antithetical effect of increase of the void-radius on CRSS. For example, at the temperature of 653 K as shown in Fig. 9, the maximum CRSS corresponds to the voids of radius 3.8 nm, whereas the mean and maximum radii are about 7.6 nm and 12.5 nm respectively. This indicates that as soon as the depinning takes place at the voids of 3.8 nm, the link-length becomes large enough to suppress the critical stress for the rest of the obstacles. This behaviour of occurrence of the maximum hardening at smaller voids has been observed for all the irradiation temperatures and the results are summarized in Table 2. The second column of Table 2 shows the void-sizes ($\hat{r}$) corresponding to maximum hardening, with respect to the mean values of the size distribution $\left(\langle r\rangle = \int_{r_{min}}^{r_{max}} r f_o(r) dr\right)$, whereas the next column represents these values in terms of the maximum radii ($r_{max}$) of nanovoids available in the size distributions. We find that this effect is more prominent at 653 K and 853 K, where the void-sizes corresponding to the peak hardening are below 15% of the maximum void-sizes. Nevertheless, for the intermediate temperatures, this ratio is close to 30%.

## 4. Conclusions

In a nutshell, we propose a new framework capable of yielding the statistics of pinning of dislocations at obstacles. This model takes into account spatially homogeneous, three-dimensional distributions of dislocation lines and pinning defects, which can be modelled as spherical in shape. Only a minimal set of material parameters including the arbitrary size-distribution of the pinning defects is needed as the input for this framework. Application of this method has been illustrated by determining the effect of dislocation-nanovoid interactions on the climb and glide motions of pinned dislocation in type 316-stainless steel, which has been irradiated at different temperatures. It has been found that by using the statistics of pinning, in conjunction with the available knowledge of fundamental dislocation-defect interactions, some new aspects regarding the collective behaviour of dislocations are revealed, which would otherwise have been inaccessible in absence of the statistical information of pinning. Even though a particular case study has been selected here for demonstration, this approach is a generalized one and can be used for other types of defects and interactions as well. As more and more details of the intrinsic parameters of the material are available, this model can further be refined to yield better results of physical significance.

**Table 1**

The volume fraction of the nanovoids and the network-dislocation density in type 316-SS at different irradiation temperatures estimated using the empirical relations [26].

| Temperature (K) | Volume fraction $\eta_o$ (%) | Network-dislocation density $\rho_\perp$ (in cm$^{-2}$) |
|---|---|---|
| 653 | 1.83 | $3.09 \times 10^{8}$ |
| 713 | 4.02 | $7.13 \times 10^{10}$ |
| 763 | 3.91 | $1.00 \times 10^{11}$ |
| 803 | 3.40 | $7.12 \times 10^{10}$ |
| 853 | 2.63 | $4.63 \times 10^{10}$ |

**Table 2**

Radius of the nanovoid at which the void-induced hardening is maximum, is shown in comparison to the mean and maximum radii associated with the corresponding size-distribution.

| Temperature (K) | $\frac{\hat{r}}{\langle r \rangle}$ | $\frac{\hat{r}}{r_{max}}$ |
|---|---|---|
| 653 | 0.21 | 0.12 |
| 713 | 0.46 | 0.23 |
| 763 | 0.48 | 0.29 |
| 803 | 0.45 | 0.29 |
| 853 | 0.30 | 0.14 |

**Figure captions :**

**Fig. 1**. Unit cube of the model material showing spherical pinning defects and dislocation lines.

**Fig. 2**. (a) Schematic representation of a nanosized spherical obstacle assumed to be placed inside a cubical cell. (b) A large number of such cubical cells are required to fill the unit volume of the sample solid. Some of these cells would accommodate parts of the dislocation-network or pinning obstacles or both. The top view presents the two-dimensional illustration of the cells. A typical cube having obstacle radius $r$ is zoomed with the dislocation segment connecting any two arbitrary faces of the cube.

**Fig. 3**. Two-dimensional illustration of shifting an off-centre obstacle to place it inside the nearest cubical cell.

**Fig. 4**. A few possible arrangements of the dislocation segment inside the cubical cell. The value of $\langle L_{\perp}(r)\rangle$ involves the ensemble average of all such possibilities.

**Fig. 5**. $\frac{dN_{\perp o}}{dr}$ and $\frac{dN_{o}}{dr}$ plotted against the void radius at irradiation temperatures of (a) 653 K, (b) 713 K, (c) 763 K, (d) 803 K and (e) 853 K. At higher temperatures, the number of pinned dislocation segments exceeds that of the pinning defects, thereby indicating multiple pinning at the same site.

**Fig. 6**. Climb of a dislocation connected to two neighbouring nanovoids. If the voids are mobile enough, they can get dragged along with the climbing dislocation.

**Fig. 7**. Increments in link-length ($\lambda$) with increase in the displacement, $y$, at (a) 653 K and (b) 713 K temperatures. (c) and (d) Corresponding volume fractions (void-swelling) of the nanosized voids. As discrete bins of finite sizes have been used for representing the void size-distributions during numerical implementation of the statistical framework, step-like feature appears in the resulting plots.

**Fig. 8**. Link-length plotted with respect to the angle of curvature at (a) 803 K and (b) 853 K respectively. (c) and (d) Appreciable recovery is observed in the form of reducing volume fraction of the nanovoids at these temperatures.

**Fig. 9**. Typical trend of the normalized (w.r.t. the maximum value) CRSS plotted as a function of the corresponding size of the pinning nanovoid, with dynamically increasing link-length. Clearly, the hardening attains its maximum for a void-size ($\hat{r}$), which is much smaller than both the mean ($\langle r \rangle$) and maximum ($r_{max}$) values of the size distribution.

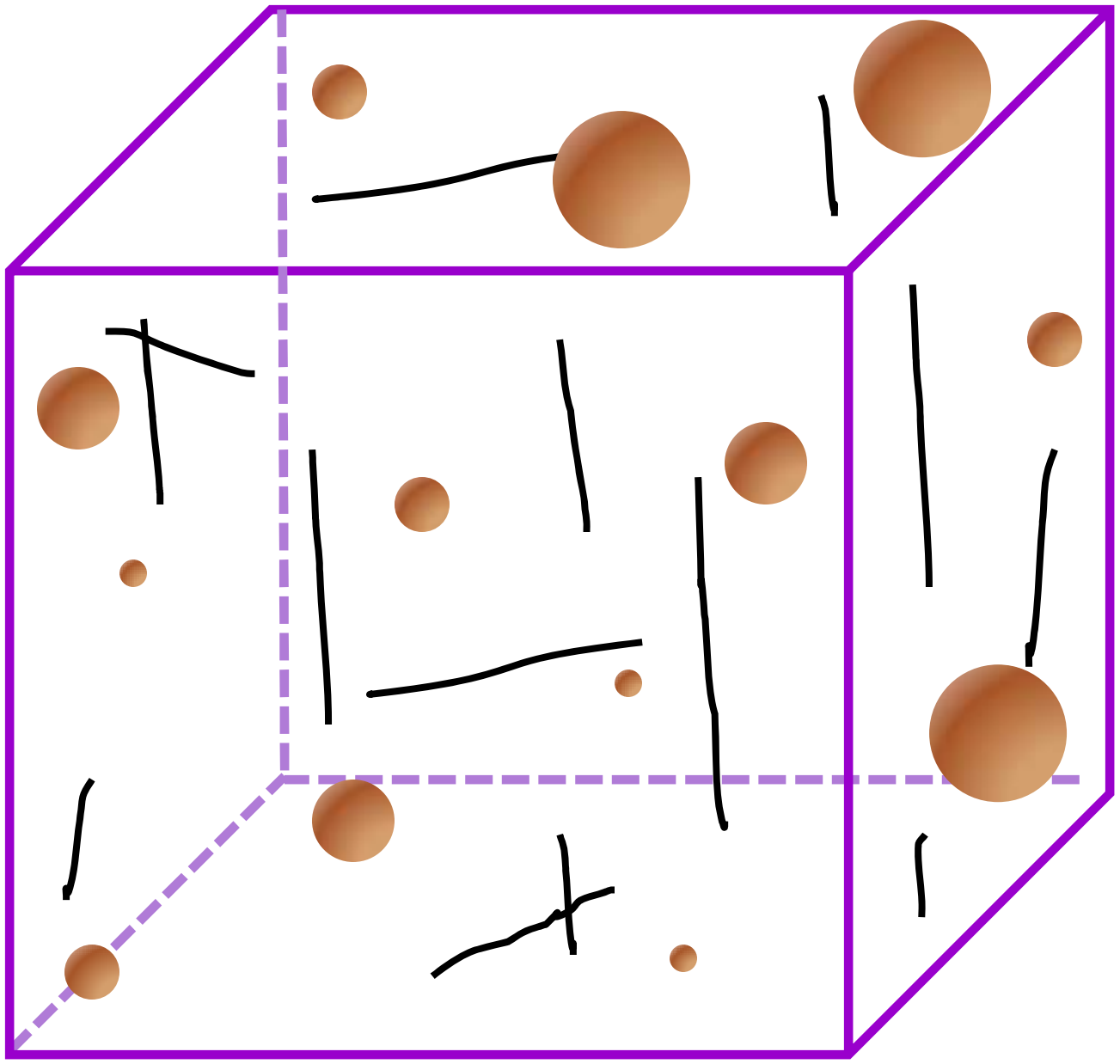

Fig. 1

Figure(s)

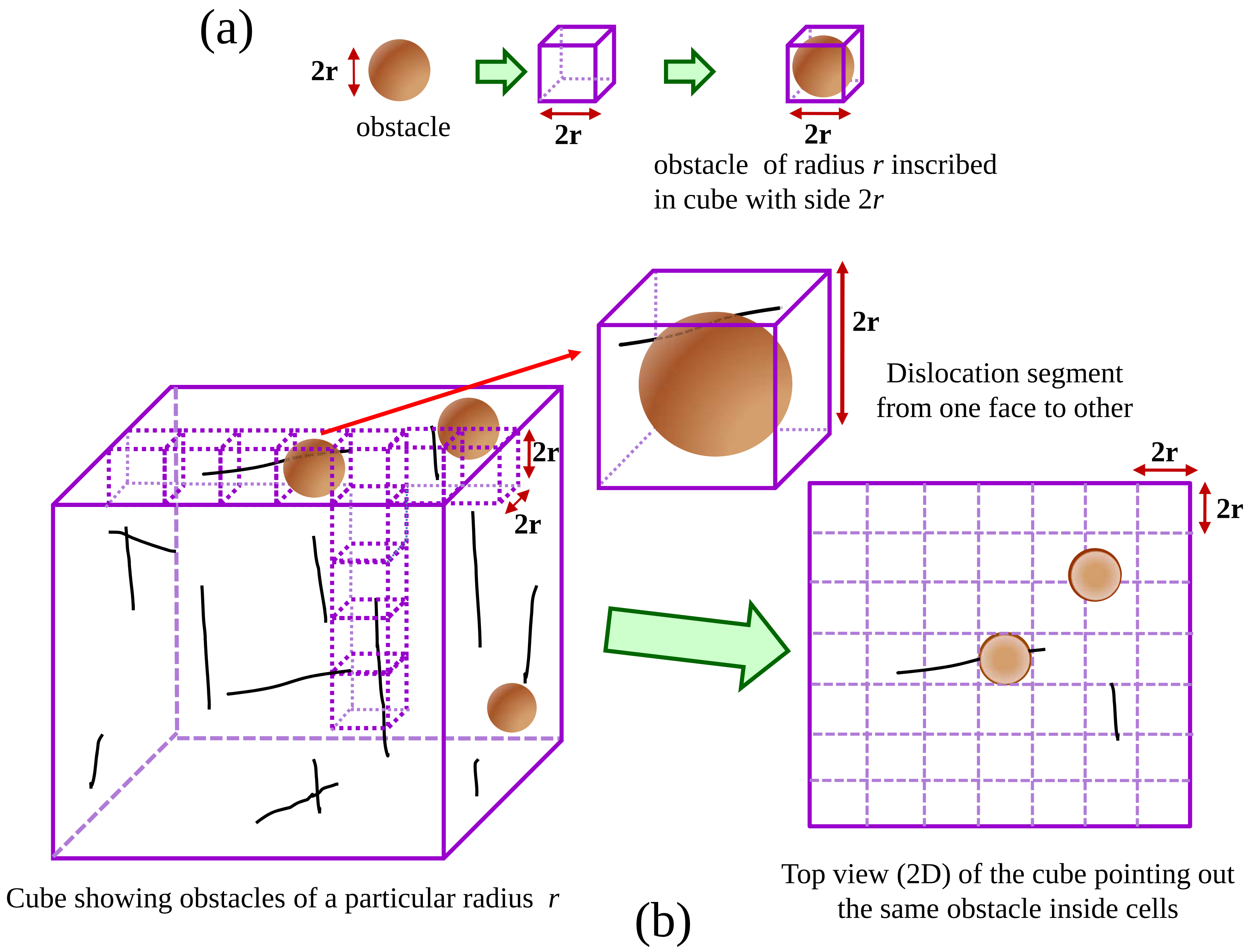

Fig. 2

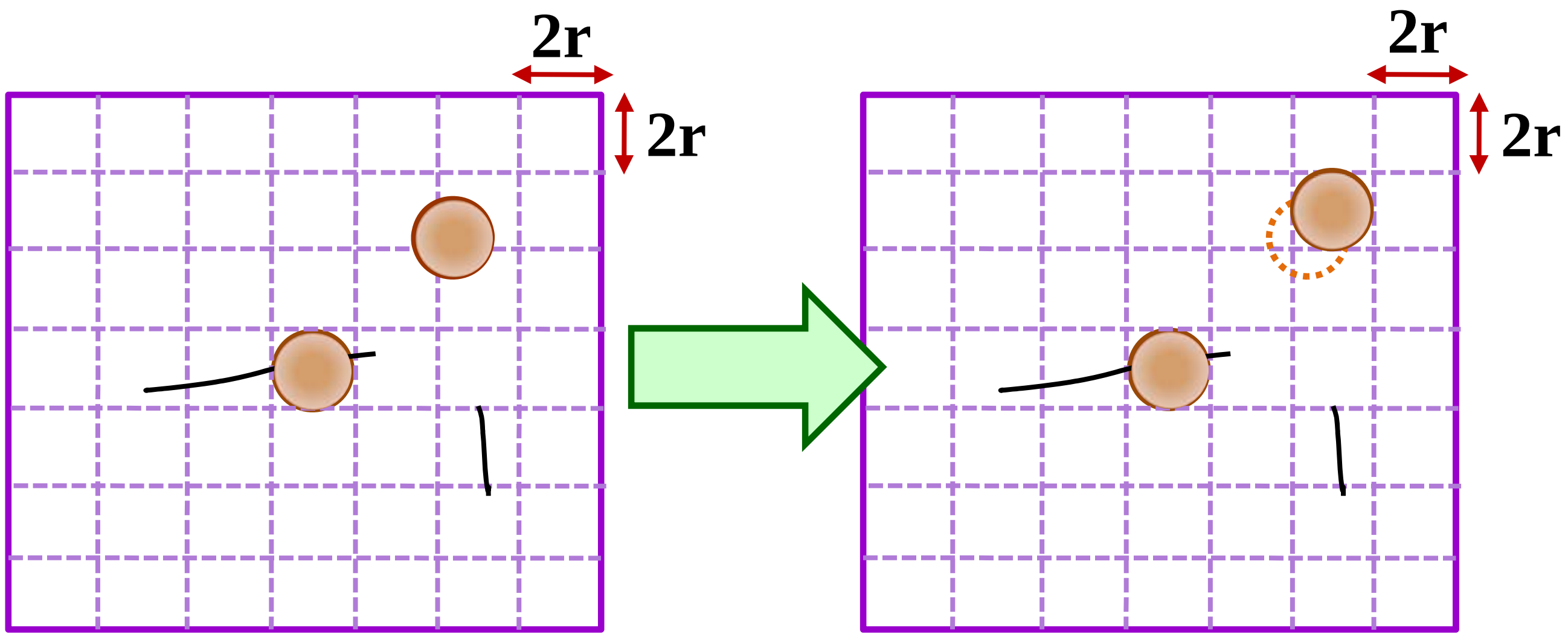

Fig. 3

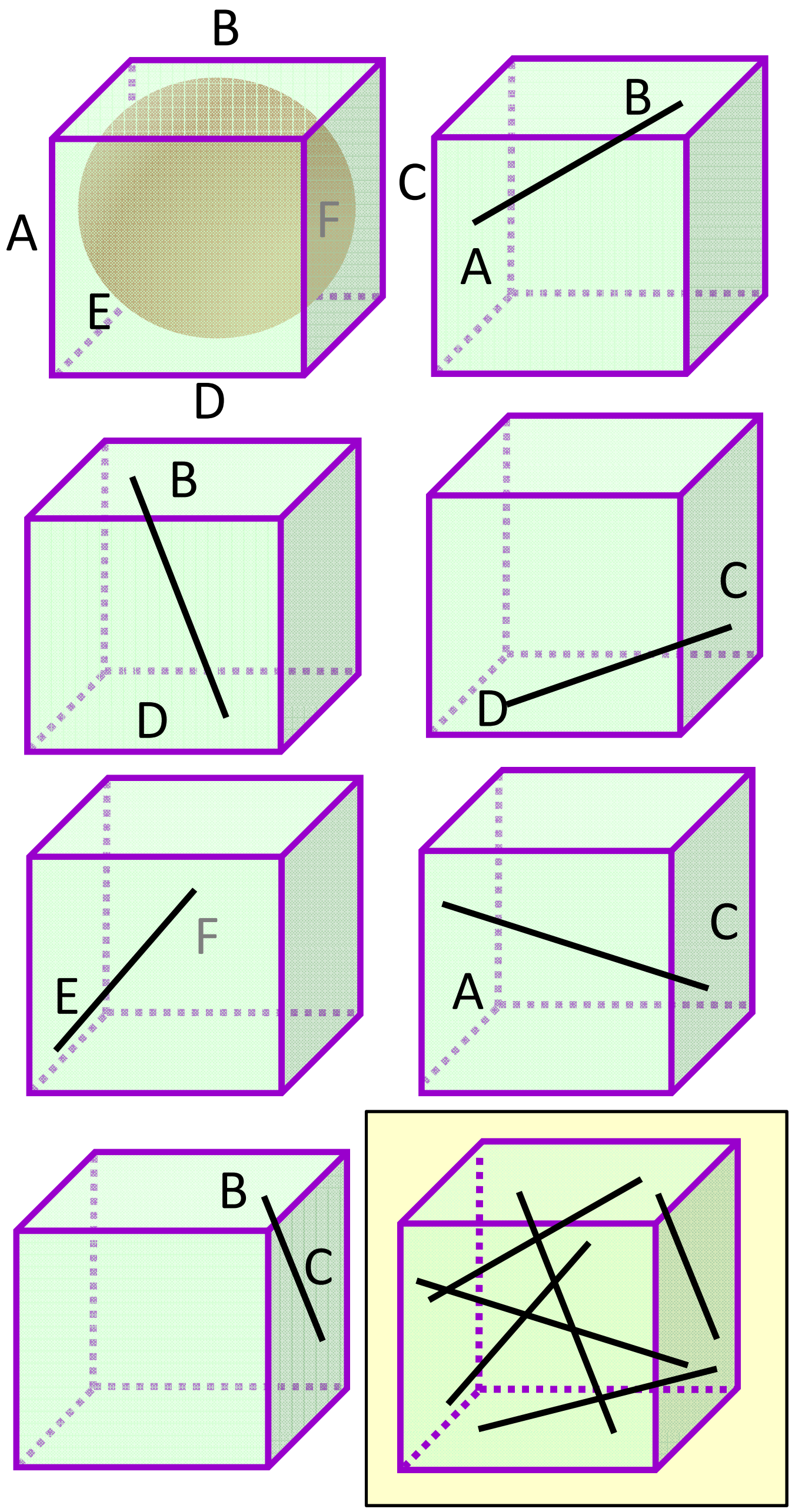

Fig. 4

Figure(s)

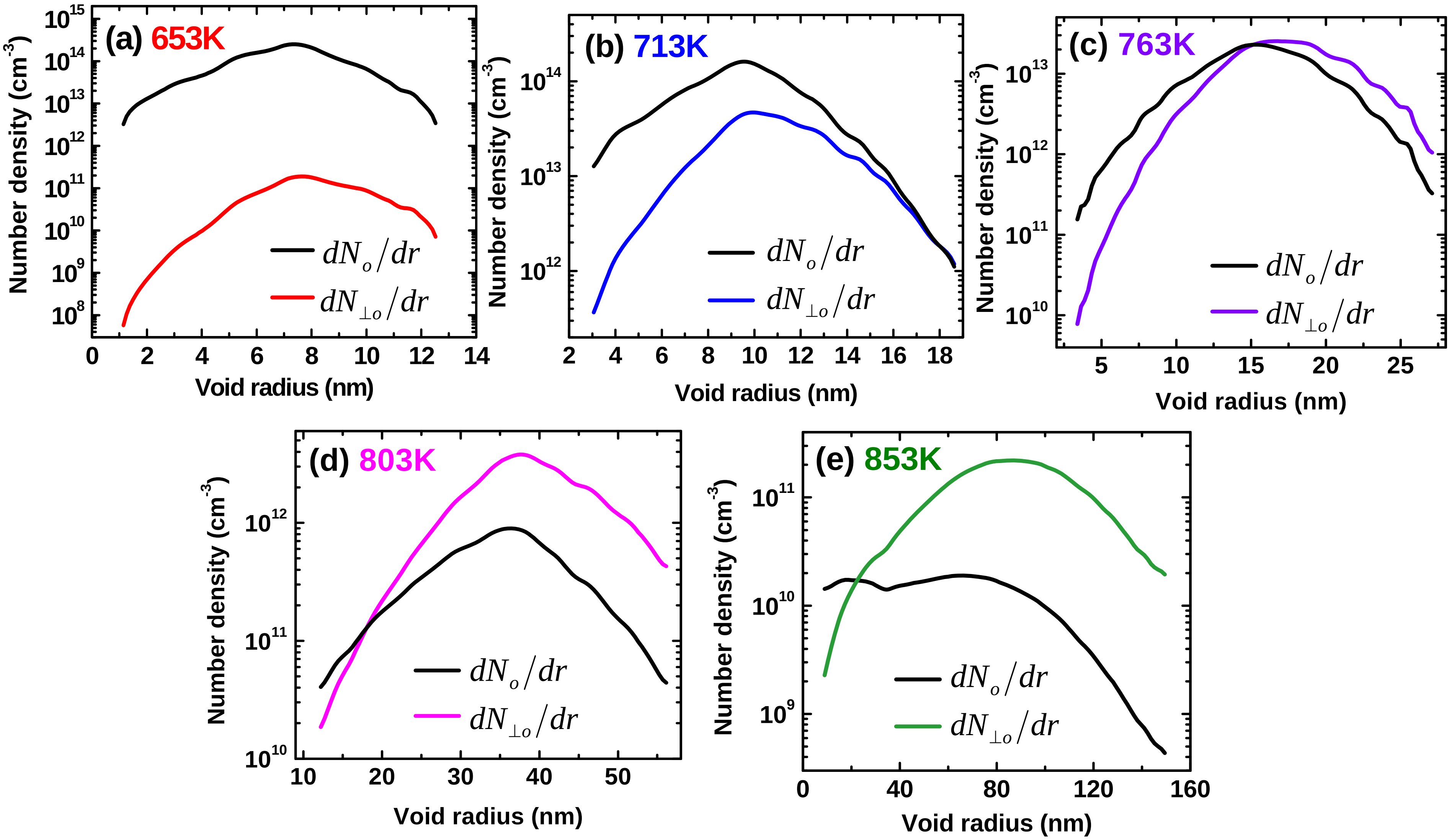

Fig 5

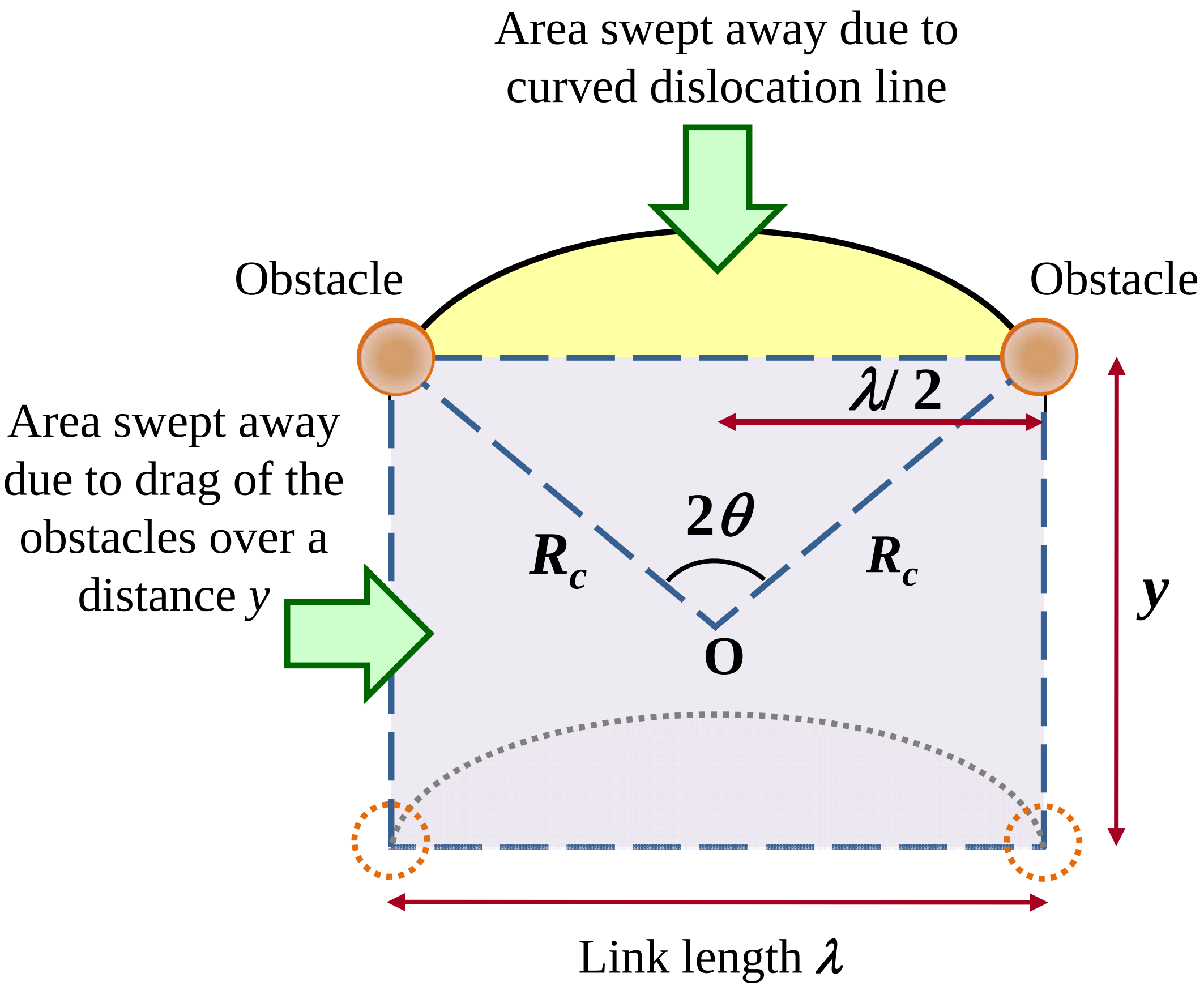

Fig. 6

Figure(s)

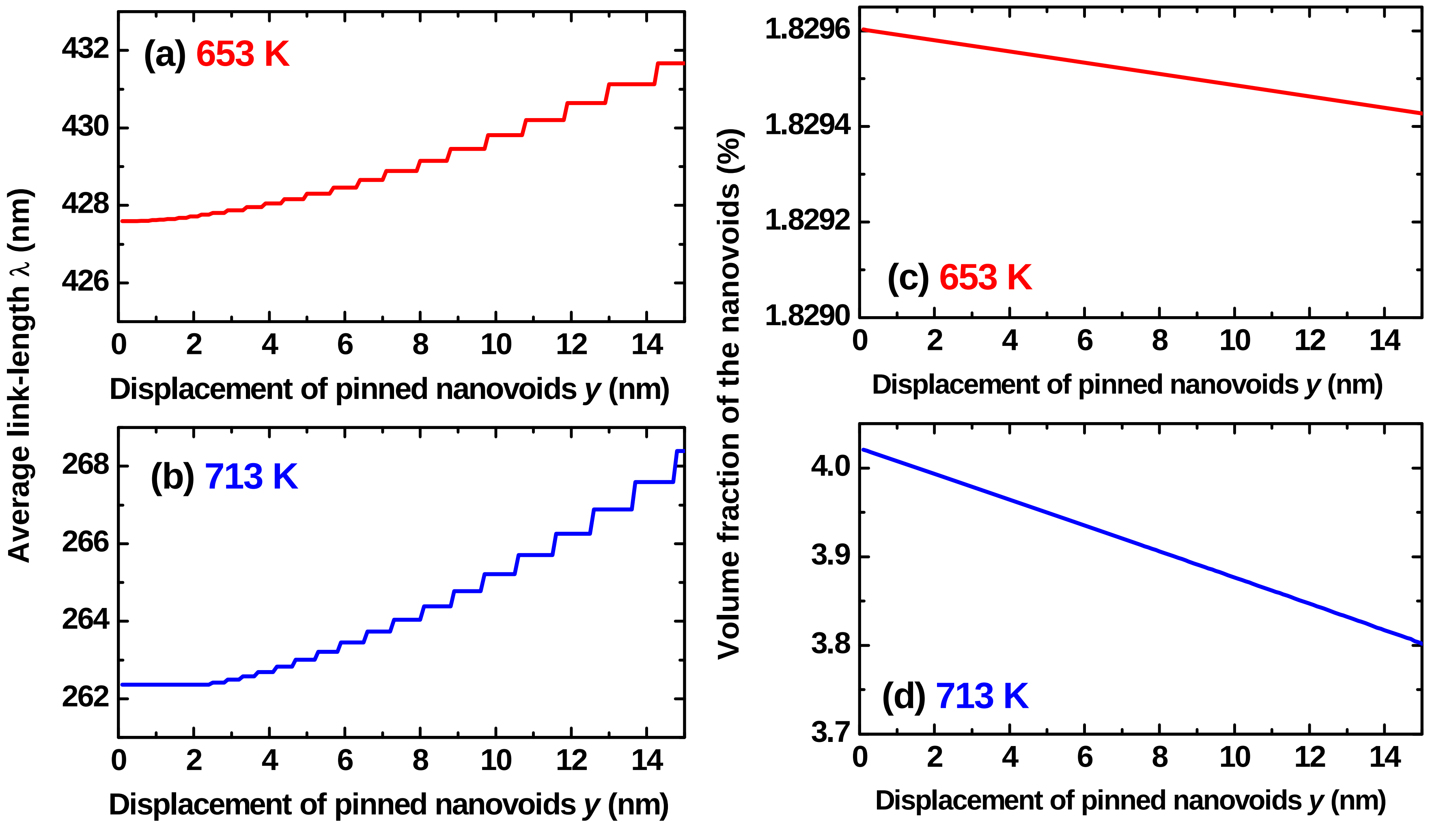

Fig. 7

Figure(s)

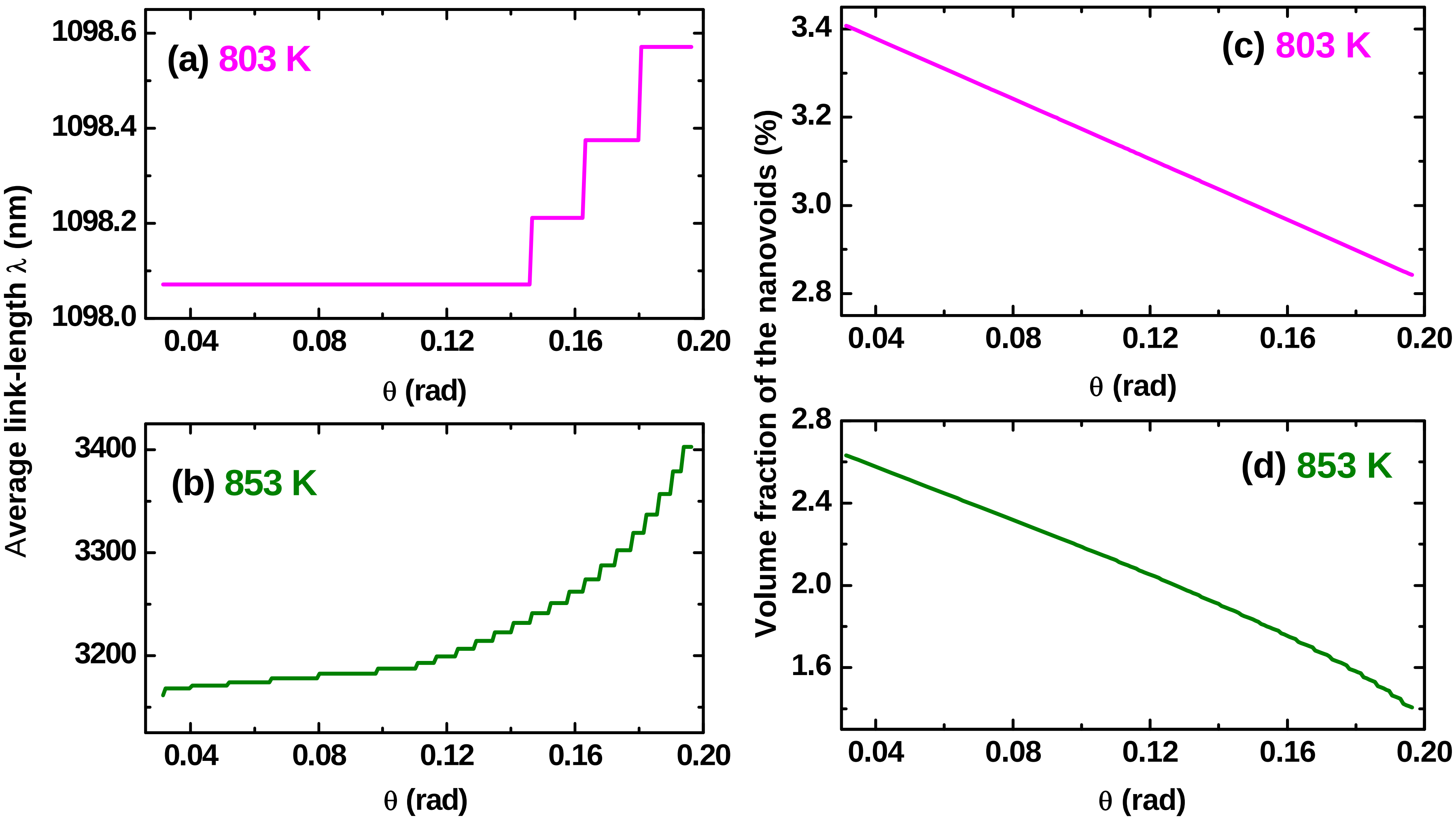

Fig. 8

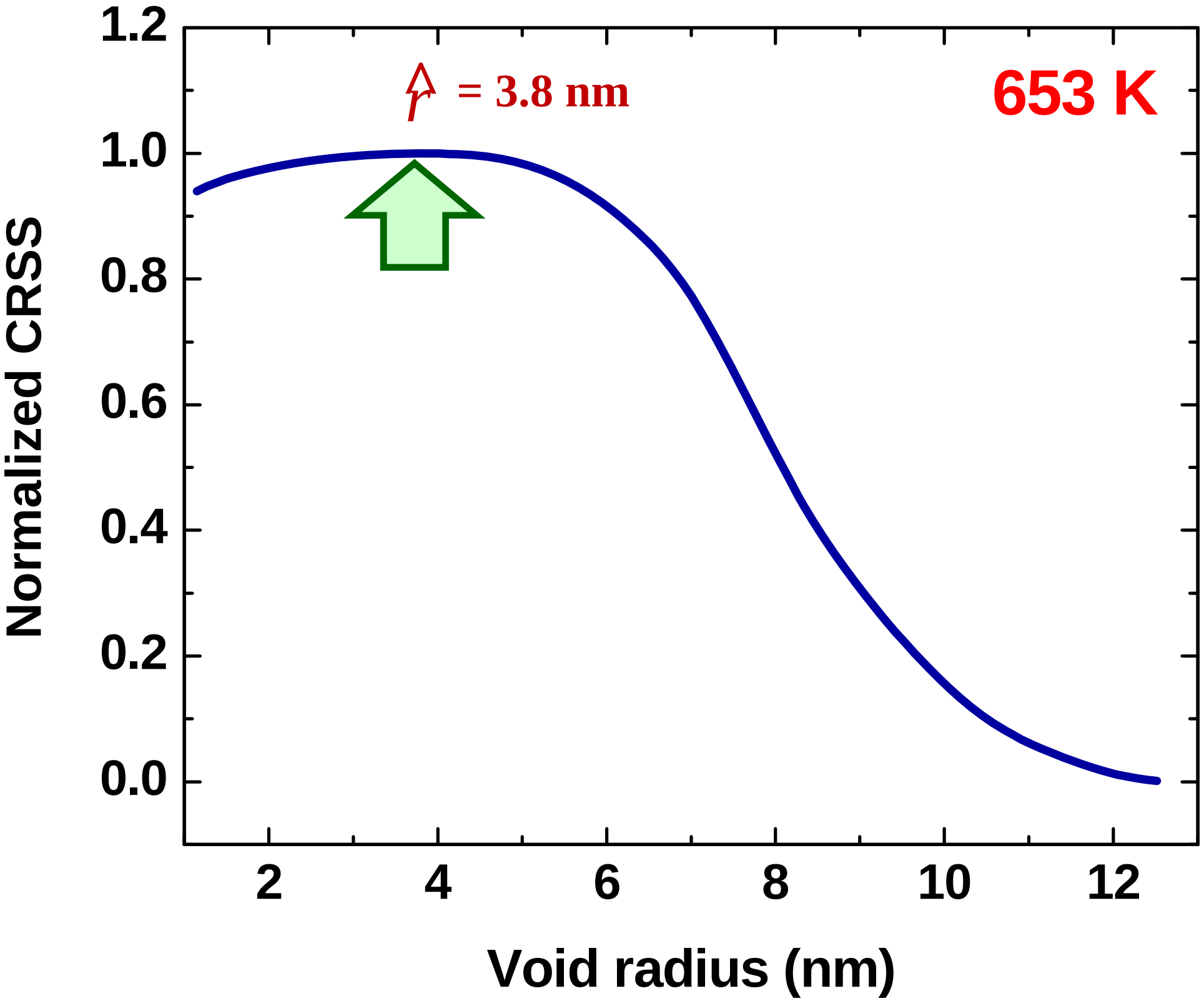

Fig. 9